\documentclass[prl,twocolumn]{revtex4}
\usepackage{graphicx}
\usepackage{psfrag}
\usepackage{psfig}

\begin{document}
\title{Asymptotic behaviour of the Rayleigh--Taylor instability}
\author{Laurent Duchemin$^1$, Christophe Josserand$^2$ and Paul Clavin$^3$\\
\small $^1$ Department of Applied Mathematics and Theoretical Physics, University of Cambridge,
Cambridge CB3 0WA, United Kingdom\\
\small  $^2$ Laboratoire de Mod\'elisation en M\'ecanique\\
\small UPMC-CNRS UMR 7607, 4 place Jussieu, 75252 Paris C\'edex 05 France\\
\small $^3$ IRPHE, Universit\'es d'Aix-Marseille I \& II-CNRS, 49 rue Joliot-Curie, BP 146, 13384 Marseille Cedex France\\
}

\begin{abstract}
We investigate long time numerical simulations of the inviscid Rayleigh-Taylor instability at Atwood 
number one using a boundary integral method. We are able to attain the asymptotic behavior for the
spikes predicted by Clavin \& Williams\cite{clavin} for which we give a simplified demonstration.  In
particular we observe that the spike's curvature evolves like $t^3$ while the overshoot in acceleration
shows a good agreement with the suggested $1/t^5$ law. Moreover, we obtain consistent results
for the prefactor coefficients of the asymptotic laws. Eventually we exhibit the self-similar behavior of
the interface profile near the spike.
\end{abstract}
\maketitle

\section{Introduction}
\begin{figure}[h]
\begin{center}
\psfrag{vertical velocity}{\Large Vertical velocity}
\psfrag{time}{\Large time}
\psfrag{-pi}{$-\pi$}
\psfrag{0}{$0$}
\psfrag{-1}{$-1$}
\psfrag{-2}{$-2$}
\psfrag{1}{$1$}
\psfrag{2}{$2$}
\psfrag{3}{$3$}
\psfrag{4}{$4$}
\psfrag{5}{$5$}
\psfrag{6}{$6$}
\psfrag{7}{$7$}
\psfrag{8}{$8$}
\psfrag{5}{$5$}
\psfrag{10}{$10$}
\psfrag{pi}{$\pi$}
\psfrag{2p}{$2\pi$}
\psfrag{3p}{$3\pi$}
\psfrag{4p}{$4\pi$}
\psfrag{5p}{$5\pi$}
\psfrag{6p}{$6\pi$}
\psfrag{6.5}{$t=6.5$}
\psfrag{7.5}{$t=7.5$}
\psfrag{9.5}{$t=9.5$}
\centerline {\includegraphics[width=6cm]{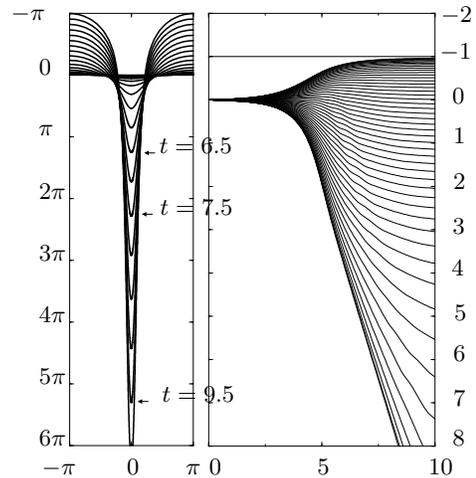}}
\end{center}
\caption{Snapshots of the interface subject to the Rayleigh-Taylor instability for time
ranging from $t=0$ to $10$, starting with a small amplitude sine mode (left). On the
right is shown the velocity of several points along the interface, non-dimensionalized with the stationnary 
bubble rising velocity $\sqrt{g/3k}$, as a function of time.\label{all}}
\label{profils}
\end{figure}

The Rayleigh-Taylor (RT) instability appears when, under gravity, an heavy liquid is placed over a lighter one\cite{rayleigh}. 
This instability is crucial for our understanding of different phenomena in fluid mechanics: mixing, thermal 
convection (\cite{kada} and cited ref. herein) and also finger number selection in splashes\cite{gueyffier}. 
It is also important in inertial confinment fusion where the mass ablation provides a stabilizing effect to
the interface instability\cite{sanz}. Without ablation, after the exponential growth of the perturbations due
to the linear RT instability, nonlinear profiles develop through the formation of bubbles of lighter fluid 
rising into the heavier one and falling
 spikes of the heavier liquid penetrating the lighter one. In the 
general situations of viscous fluids which are immiscible and/or have Atwood number not equal to unity ($A_T=(\rho_h
-\rho_l)/(\rho_h+\rho_l)$ with $\rho_h$ and $\rho_l$ being the density of the heavier and lighter fluids
respectively), famous mushrooms-like structures grow for 
larger times\cite{kada,sohn1,sohn2}. The limit of an inviscid fluid above a vacuum ($A_T=1$) without surface tension plays 
a specific role since no stabilizing effects are present in the linear dynamics. Numerous theoretical
and numerical work have focused on this idealized limit in order to track insights into the instability
itself\cite{layzer,zhang,hazac,abar,mika,inog}. It has been
shown using a conformal mapping that a finite time singularity might appear in the conformal 
plane\cite{tanveer} and
it is also suspected that for some sufficiently irregular initial conditions finite time singularities should also 
be observed in the physical plane. However, starting with sufficiently smooth initial conditions, the 
asymptotic dynamics\cite{mika,zhang,inog}  presents a constant velocity rising bubble separated by free falling tiny spikes as 
displayed on figure \ref{profils}. Although the
rising bubble motion has been described using local properties of the flow\cite{gonch}, the asymptotic dynamics
of the spikes is far from being well understood. The single mode approach gives a fair 
description of the constant velocity of the rising bubble ($v_b=\sqrt{g/(3k)}$ where $g$ is the 
acceleration of the gravity and $k$ the wavenumber of the perturbation) but gives only partial results
for the spike\cite{zhang}. The fluid there obeys free fall dynamics to a good approximation and the 
pressure field of the flow leads to an overshoot in the acceleration. The accelerated
motion of the liquid stretches the spike geometry and one expects self-similar behaviour of the
tip of the spikes. 
Recently, an asymptotic theory using a parallel flow description of the velocity field near the 
spikes has been constructed \cite{clavin}. The interface dynamics is nonlinear for large time and can 
be described using the theory of characteristics which gives rise to finite time singularity 
solutions. In the case of regular dynamics a self-similar 
description of the peak is obtained for large time: the maximal curvature of the interface at the
peak tip is found to behave like the cubic power of time $t^3$. Moreover, the spike position,
following the free fall $ \frac12 g t^2$ at leading order, is shown to converge to the constant 
acceleration $g$ with an overshoot in acceleration decreasing like $t^{-5}$. In this letter, we 
present a numerical study of the Rayleigh-Taylor instability which focuses on the large time dynamics 
of the spikes in order to investigate the self similar dynamics predicted in \cite{clavin}.
We consider  the dynamics for an inviscid liquid (heavy)  with an exterior fluid of zero density 
($A_t=1$) and no surface tension. The numerics use a boundary integral method (BIM later on).
Due to strong numerical instabilities, a careful treatment of the
interface using conformal mapping is needed as explained below. The results are then 
shown and compared with the theory.

\section{Asymptotic analysis and numerical method}

\begin{figure}[h]
\psfrag{omega}{$\Omega$}
\psfrag{zeta}{$\zeta=f(z)$}
\psfrag{x}{$y$}
\psfrag{y}{$x$}
\psfrag{function}{$f(z) = e^{-i z}$}
\psfrag{M}{$M$}
\centerline {\includegraphics[width=8cm]{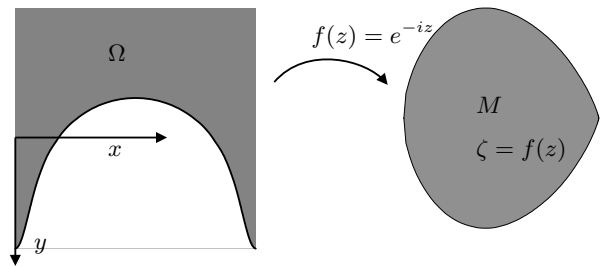}}
\caption{\label{map} Conformal map used to transform the physical periodic plane $\Omega$ into a closed domain $M$.}
\end{figure}

We consider the two-dimensional motion of an inviscid fluid 
above a vacuum, subject to a negative acceleration $-g$. A periodic
sine perturbation of the interface of wave number $k$ is implemented as initial conditions.
Neglecting surface tension, the equations of motion have no 
 control parameter after rescaling the time, the position and the velocity potential $\varphi$ by 
 factors $\sqrt{gk}$, $k$ and $\sqrt{k^3/g}$ respectively.
The interface is described by $y=\alpha(x,t)$, where $y$ is the direction along the 
gravity and $x$ orthogonal to it (see figure \ref{map}). 
The velocity field ${\bf U}=(u,v)$ satisfies the dimensionless Euler equation
$$ \frac{d{\bf U}}{dt}=-{\bf \nabla}P +{\bf e_y} $$
where $P(x,y,t)$ is the pressure, ${\bf e_y}$ the non-dimensional acceleration due to gravity and 
the fluid density $\rho=1$.
 The kinetic equation for the interface reads~:
$$ \frac{\partial \alpha(x,t)}{\partial t}+u\frac{\partial \alpha(x,t)}{\partial x}=v$$
with the velocity field $(u,v)$ evaluated at the interface $(x,\alpha(x,t))$. 
Starting at time $t=0$ with a small sine amplitude interface, we observe for large time 
that the fluid particles located in the vicinity of the tiny spikes come from an almost free fall from the 
initial interface region. Therefore, following \cite{clavin}, 
we assume quasi-parallel steady flow for the velocity field which gives then in the tip region $ |u|\ll |v|$ and~:
$$ v \sim \sqrt{2y}$$ 
with $y \sim \frac12 t^2$ for large time.
Writing a perturbation expansion of the velocity field in the tip region $|x| \ll y$, we in fact consider:
$$ v=\sqrt{2(y+f(x,y,t))} $$
with $ f(x,y,t) \ll y$. Taking a Taylor expansion in $x$ of the perturbation $f$, we obtain by symmetry:
$$ v = \sqrt{2y}+\frac{f_0(y,t)}{\sqrt{2y}}+\frac{x^2}{2} \frac{f_2(y,t)}{\sqrt{2y}} + O(x^4) $$
We limit our expansion to the second order in $x$ for the velocity field later on.
Incompressibility gives~:
$$ u=-\left(\sqrt{\frac{1}{2y}}+\frac{\partial (f_0(y,t)/\sqrt{2y})}{\partial y}\right)x + O(x^3).$$
At the leading order (where we neglect even the perturbation $f(x,y,t)$) we obtain the following 
expression for the interface location:
$$  \frac{\partial \alpha(x,t)}{\partial t}-\frac{x}{\sqrt{2\alpha(x,t)}} \frac{\partial \alpha(x,t)}{\partial x}=\sqrt{2\alpha(x,t)} $$
which can be solved using the methods of charasteristics (see \cite{clavin}).
Writing $\alpha(x,t)=t(\frac{t}{2} - \gamma(x,t))$ and noting that $\gamma(x,t)\ll t/2$ in the spike region, 
we obtain, after linearisation~:
$$ \frac{\partial \gamma(x,t)}{\partial t}-\frac{x}{t}\frac{\partial \gamma(x,t)}{\partial x}=0 $$
which has self-similar solution of the form $ \gamma(x,t)=\theta(xt)$. A first conclusion can be drawn 
about the curvature of the interface at the tip,  $\kappa=-\partial^2\alpha/\partial x^2|_{x=0}$, which is 
thus found to increase as the cubic power of time~:
\begin{equation}
 \kappa=t^3 \theta''(0)
\label{cubic}
\end{equation}

The next order terms of the expansion allow the determination of the function $f_0(y,t)$ near the tip.
Using the constant value of the pressure at the interface we use the projection of the Euler
equation at the interface on its local tangent~:
$$ \frac{du}{dt} + \frac{\partial \alpha(x,t)}{\partial x} \frac{dv}{dt} = \frac{\partial \alpha(x,t)}{\partial x}. $$
Since on the interface $dP(x,\alpha(x,t),t)/dx=0$.
We develop this equation at first non-zero order (which will end up to be the first order in $x$) 
with the expansion $\theta(xt)=\theta(0)+x^2t^2 \theta''(0)/2+O(x^4)$. 
Remembering that $|f| \ll y$, we can neglect also the larsge scale terms $\partial^2 (f_0(y,t)/\sqrt{2y})/\partial t \partial y$ and $\sqrt{2y}\partial^2 (f_0(y,t)/\sqrt{2y})/\partial y^2$ with respect to the others.
We obtain finally for the tip position $y=y_s$~:
$$ \frac{\partial f_0(y_s,t)}{\partial t} +\sqrt{2y_s}\frac{\partial f_0(y_s,t)}{\partial y}=\frac{df_0(y_s,t)}{dt}=
\sqrt{\frac{2}{y_s}} \frac{1}{\kappa} $$

Recalling that:
$ \frac{dy_s}{dt}=\sqrt{2y_s}+\frac{f_0(y_s,t)}{\sqrt{2y_s}} $
we obtain for the tip acceleration at leading order:
\begin{equation}
 \frac{d^2y_s}{dt^2}=1+\frac{1}{\sqrt{2 y_s}} \frac{df_0(y_s,t)}{dt}=1+\frac{2}{t^5 \theta''(0)} 
\label{quint}
\end{equation}
which corresponds to an overshoot in the spike acceleration decreasing as the fifth
power of time.

The numerical method is elaborated using the incompressible and potential properties
of the flow. The velocity field can thus be evaluated everywhere when the velocity 
potential is known on the interface thanks to Cauchy's theorem, in the spirit of pionnering works\cite{vinje,cokelet,BMO,MZ}. 
The non-dimensional Bernoulli equation on the free surface reads~:
\begin{equation}
\frac{\partial \varphi}{\partial t} = -\frac{1}{2}(\nabla \varphi)^2 + y,
\end{equation}
where the velocity potential $\varphi$ is a harmonic function in the 
fluid domain $\Omega$~:
\begin{equation}
\Delta \varphi = 0
\label{lap}
\end{equation}

The kinematic condition on the free surface expresses the fact that 
fluid particles move with the same normal velocity than the free surface 
itself~:
\begin{equation}
\frac{d \bf x}{dt} \cdot {\bf n} = \nabla \varphi \cdot {\bf n}
\label{kin}
\end{equation}

Knowing $\varphi$ on the free surface at a given time-step, 
we search for the solution of equation (\ref{lap}) that satisfies 
this boundary condition (\ref{kin}). We use the complex potential $\beta(z) = \varphi + i \psi$ 
and the conformal map $f(z) = exp(-i z)$ (Cf. Figure \ref{map}), 
where $z=x+iy$ and $\psi$ is the stream function. The conformal map transforms the periodic 
domain $\Omega$ into the closed domain $M$. 
Since $\psi$ is harmonic inside $\Omega$, $\beta(z)$ is analytic 
inside $\Omega$ and therefore $\gamma(\zeta)=\beta(f(z))$ 
is analytic inside $M$. Using Cauchy's theorem, we obtain a
Fredholm equation of the second kind for the stream function $\psi$ which is solved using
discretization of the free surface ($\partial \Omega$ and thus $\partial M$).
This linear system of equations is solved using a $LU$ decomposition. 
Once we know $\psi$ on each point on $\partial M$, the complex velocity of each marker in the physical plane is given by~:
\begin{equation}
\frac{d \beta}{dz} = u - i \, v
\end{equation}
where $u$ and $v$ are the horizontal and vertical velocities 
respectively. 
This complex velocity is computed with a finite difference scheme 
using the values of the complex potential on the collocation points 
on $\partial \Omega$.
The position of the surface markers (kinematic condition) 
and the value of the velocity potential on each of these markers 
(Bernoulli equation) are then updated in time using a fourth 
order Runge-Kutta method. 

\section{Results and discussions}
We have performed numerical simulations of the Rayleigh-Taylor instability using the numerical
method described above. We start with a small amplitude sine-mode. The unavoidable
numerical noise cannot be damped by the numerics and the calculations always end up subject to 
numerical instabilities. 
Nevertheless, we emphasize that the numerical scheme used here is remarkably robust and 
can be accurately evolved to reach the large time where the scalings predicted by the theory \cite{clavin} are valid.
 Comparing our simulations with recent numerical works\cite{sohn1,sohn2,hazac}, we have been able to run the
 dynamics at least twice as far which corresponds roughly to an increase of a factor of $8$ in the tip's curvature.

\begin{figure}[h]
\centerline {
\includegraphics[width=7cm]{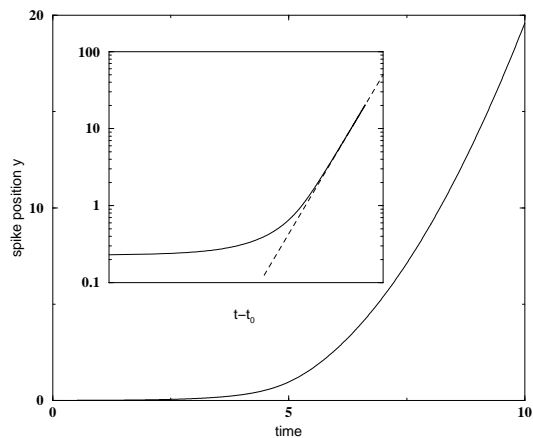}}
\caption{Position of the spike $y_s(t)$ as a function of time. The inset shows in a log-log plot of the spike
position (black curve) as function of time $t-t_0$ with $t_0=3.74$ obtained by a second order 
polynomial fit of $y_s$. the dashed line shows the expected behavior $\frac12 t^2$. }
\label{spike}
\end{figure}

\begin{figure}[h]
\centerline {
\includegraphics[width=7cm]{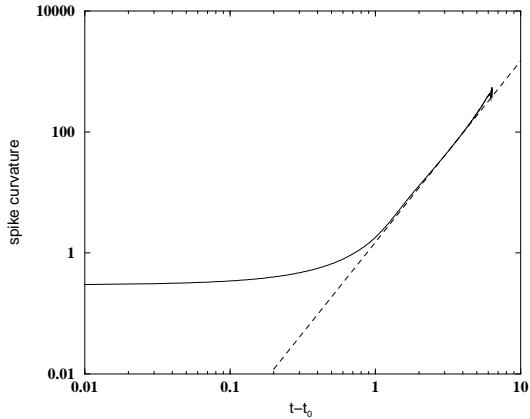}}
\caption{Spike curvature $\kappa_s$ calculated at the tip $y=y_s$ as function of the delayed time $t-t_0$ 
in a log-log plot. The dashed line displays the cubic law (\ref{cubic}) with $\theta''(0)=1.5$.}
\label{courbure}
\end{figure}

\begin{figure}[h]
\centerline {
\includegraphics[width=7cm]{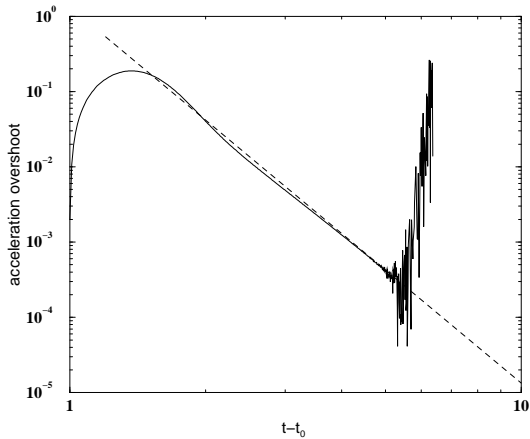}}
\caption{Overshoot in acceleration, defined as the difference between the tip acceleration and the
gravity. The plot is in log-log scale and with the delayed time $t-t_0$. The dashed line shows the 
theoretical prediction (\ref{quint}) using the value of $\theta''(0)$ obtained from figure \ref{courbure}}
\label{accel}
\end{figure}

\begin{figure}[h]
\centerline {
\includegraphics[width=6cm]{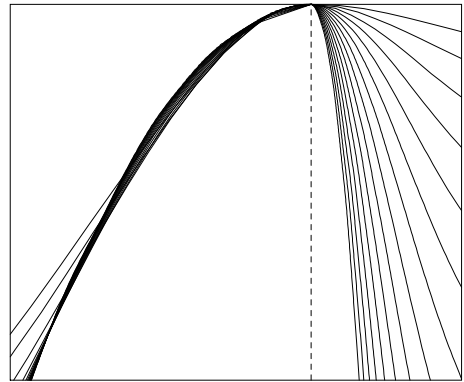}}
\caption{Self-similar structure of the tip: the interface profile around the spike have been superimposed 
on the right side of the figure for different time $t$ ranging from $4$ to $12$. The left side of the figure
shows the same curves rescaled by factor $1/(t-t_0)$ and $(t-t_0)$ for the $x$ and $y$ coordinates
respectively, following the scaling behavior predicted by the theory.}
\label{similar}
\end{figure}

The position of the spike is shown on figure \ref{spike} as function of time. We observe that the 
asymptotics dynamics are very well approximated by the relation $ y_s =\frac12 g(t-t_0)^2$ as shown in 
the inset to the figure with $t_0=3.74$. This remarkable behavior, in good agreement with the free fall 
 hypothesis,
suggests that $t_0$ is the time delay accounting for the initial exponential development of the instability. 
We will therefore present further data on the curvature dependance and the acceleration of the tip as functions
of this delayed time $t-t_0$ instead of $t$.
The curvature $\kappa_s$ at the tip is then shown on figure \ref{courbure}. The large time asymptotic 
behavior is similarly found to follow the cubic law (see equation \ref{cubic}) with $\theta''(0)=1.5$.
In addition, the acceleration of the tip is computed by finite differences on the tip velocity and
the overshoot in the acceleration is presented on figure \ref{accel}. We observe that the results 
look noisier than the two previous ones. Two factors can explain such noise: firstly, we are
looking to a finite difference which decreases to zero so that the numerical errors are relatively more
important. However,
we note that the overshoot in acceleration shows a good agreement with the $1/t^5$ law, noting
that no adjustable parameter is used in this comparison.
Moreover, the self similar structure of the interface near the tip has been exhibited on figure \ref{similar}.
We observe after the proper rescaling on the left part of the figure that the interface 
profiles collapse onto a single curve near the spike.

We have thus exhibited large times numerical simulations of the Rayleigh-Taylor instability which 
present asymptotic scaling behavior in agreement with theoretical predictions using Taylor
expansions of the free fall velocity field at the spike\cite{clavin}. Although our numerics always stops
due to numerical instability, we have been able to reach large time enough to exhibit the cubic power in time dependance for the spike 
curvature and the inverse of the quintinc power of time decreasing of the overshoot in acceleration.

It is our pleasure to thank J. Ashmore for useful comments.
We acknowledge also the support of CEA through the contract CEA/DIF N° 4600051147/P6H29.

\end{document}